\documentclass{iopart}
\usepackage[dvips]{graphicx}

\begin{document}

\title{Schr\"odinger cats and their power for quantum information processing} 

\author{A. Gilchrist\dag
\footnote[7]{alexei@physics.uq.edu.au}, Kae Nemoto\P, W. J. Munro\S\, T. C. Ralph\dag\, 
S. Glancy$^+$,
Samuel. L. Braunstein\ddag\ and G. J. Milburn\dag}

\address{\dag\ Centre for Quantum Computer Technology,
University of Queensland, QLD 4072, Australia }
\address{\P\ National Institute of Informatics,
   2-1-2 Hitotsubashi, Chiyoda-ku, Tokyo 101-8430, Japan.}
\address{$^+$\  Department of Physics, University of Notre Dame,
Notre Dame, Indiana 46556, USA}
\address{\ddag\  Computer Science, York University, York YO10 5DD, UK}
\address{\S\  Hewlett Packard Laboratories, Filton Road, Stoke Gifford,
Bristol BS34 8QZ, U.K}

%%%%%%%%%%%%%%%%%%%%%%%%%%%%%%%%%%%%%%%%%%%%%%%%%%%%%%%%%%%%% 
\begin{abstract}
  We outline a toolbox comprised of passive optical elements, single
  photon detection and superpositions of coherent states
  (Schr\"odinger cat states). Such a toolbox is a powerful collection
  of primitives for quantum information processing tasks. We
  illustrate its use by outlining a proposal for universal quantum
  computation.  We utilize this toolbox for quantum metrology
  applications, for instance weak force measurements and precise phase
  estimation. We show in both these cases that a sensitivity at the
  Heisenberg limit is achievable.
\end{abstract}



\maketitle

\section{Introduction}
\label{sect:intro} 
Quantum optics has played a major role in the testing of fundamental
properties of quantum mechanics and more recently in implementing
simple quantum information protocols \cite{simple,nielsen}. This has
been made possible because photons are easily produced and
manipulated.  This is especially true as the electromagnetic
environment at optical frequencies can be regarded as a vacuum and is
relatively decoherence free.

One of the earliest proposals for implementing a quantum logic gate
was made by Milburn \cite{mil88} and was based on encoding each qubit
in two optical modes, each containing exactly one photon. This was a
very elegant proposal, but unfortunately required massive and
reversible non-linearities. Such reversible non-linearities are well
beyond those presently available and hence it was thought quantum
optics would not provide a practical path to efficient and scalable
quantum computation.  Knill, Laflamme and Milburn \cite{KLM} recently
challenged this orthodoxy when they showed that given appropriate
single photon sources and detectors, linear optics alone could create
a non-deterministic two qubit gates.
% (see figure~\ref{lo}). 
%\begin{figure}[!htb]
%\begin{center}
%        \includegraphics[scale=0.5]{linear-optics}
%\end{center}
%\caption{Schematic circuit for the KLM CNOT gate. The boxes in the center of the circuit are 
%nonlinear sign shifts gates that perform the transformation $c_0 |0\rangle +c_1 |1\rangle+c_2 |2\rangle$ to 
%$c_0 |0\rangle +c_1 |1\rangle-c_2 |2\rangle$. Such a transformation can be performed using 
%on linear optical elements, single photon sources and single photon number resolving detectors. }
%\label{lo}
%\end{figure}
Furthermore they showed that near deterministic gates could be 
created from these non-deterministic gates through a 
technique of teleporting gates \cite{pre}. This therefore provided a route for
efficient and scalable quantum computation with only 
single photon sources, photon counting and linear optics.

This does however raise the question whether there are other
architectures based on different encoding schemes which have similar
characteristics.  These other architectures may have advantages in
that their optical circuits are less complex. We could trade off the
complexity of the circuit in the KLM scheme for more complicated
initial resources, for instance continuous variable multi-photon
fields. The idea of encoding quantum information on continuous
variables of multi-photon fields has emerged recently \cite{teleport}
and a number of schemes have been proposed for realizing quantum
computation in this way \cite{lloyd,sanders,kim01}. A significant
drawback of these proposals is that hard non-linear interactions
are required {\it in-line\/} of the computation and make such proposals
difficult to implement in practice.  
In contrast, a recent proposal \cite{ralph03,ralph} details a scheme for
quantum computation where the hard nonlinear interactions are only 
required for the \emph{off-line} preparation of resource states.
A required resource for this scheme is superpositions of coherent 
states (Schr\"odinger cat states). 

In this paper we outline a toolbox of techniques and states necessary
for universal quantum computation with coherent states.  This toolbox
can also be used for quantum metrology applications and we will
examine two specific examples: the detection of weak tidal forces due
to gravitational radiation \cite{simple,caves,munro} and improving the
sensitivity of Ramsey fringe interferometry \cite{bollinger,huelga}.
The paper is structured as follows: Section~(\ref{sect:toolbox})
describes the components of the toolbox, while
section~(\ref{sect:gates}) describes how to achieve a universal set of
gates sufficient for quantum information processing. Finally
Section~(\ref{sect:metrology}) illustrates two quantum metrology
examples.

\section{The Toolbox}
\label{sect:toolbox}

The base components that our toolbox will contain will be passive
linear optical elements such as beam-splitters and phase shifters. The beam-splitter 
interaction is given by
\begin{equation}
B(\theta)=\exp[i {\theta} (a b^{\dagger}+a^{\dagger} b)].
\end{equation}
Here $a$ and $b$ are the usual boson annihilation operators for the
two electromagnetic field modes at the beam-splitter. The action of
the beam-splitter is such that two coherent states $|\gamma
\rangle_{a}$ and $|\beta \rangle_{b}$ get transformed as
\begin{eqnarray}
B(\theta) |\gamma \rangle_{a} |\beta \rangle_{b}=|\gamma \cos {{\theta}}+i
\beta \sin {{\theta}}\rangle_{a} |\beta \cos {{\theta}}+
i  \gamma \sin {{\theta}} \rangle_{b}
  \label{Ho}
\end{eqnarray}
A phase shifter is just a delay with respect to the local oscillator
and can be described by the operator $P(\theta)=\exp [i \theta \hat
a^{\dagger} \hat a ]$ which just introduces a phase to the coherent
state: $P(\theta)|\alpha\rangle=|e^{i\theta}\alpha\rangle$.
From these basic components, we can construct other operators, for
instance, the displacement operator:
\begin{equation}
D(\alpha)=\exp(\alpha a^\dagger -\alpha^* a) 
\end{equation}
acting on a state $|\phi\rangle$ can be constructed by mixing that state
with a strong coherent state on a weak beam-splitter. On coherent states
the displacement operators just displace the coherent state: 
$D(\beta)|\alpha\rangle=\exp[(\beta\alpha^*-\beta^*\alpha)/2]|\alpha+\beta\rangle$.

To these passive elements we also want to add single photon counters,
which can resolve the `quanta' in the electromagnetic field, and
homodyne detectors.  While high-efficiency homodyne detection is currently
achievable \cite{Polzik Carry and Kimble 92}, single photon counters are extremely
challenging but there is an active research program to construct them
(see for instance \cite{99kim902,99takeuchi1063}).

Finally, to this collection of elements we add the ability to generate
optical ``Schr\"odinger cat'' states. These are states which are
coherent superpositions of coherent states $|\alpha\rangle$ for
different $\alpha$. In particular, we are interested in the even and
odd cat states:
\begin{eqnarray}\label{eqn:cat}
|\Psi_\pm\rangle&=& \frac{1}{\sqrt{\cal{N}_\pm}} \left[|\alpha\rangle \pm 
|-\alpha\rangle\right] ,
\end{eqnarray}
where ${\cal N}_\pm=2\pm 2 e^{-2 |\alpha|^2}$. There are several
proposals on how to generate these states (e.g. \cite{dakna97,song}).  It is
easy to show that the even (odd) cat states have only even (odd)
photon number terms --- which is where they get their name.  From this
we can see that the two states are orthogonal and a single photon
counter will be able to distinguish between them.

One of the most powerful features of this toolbox that is not
immediately obvious is that we now have the ability to easily generate
entangled states \cite{van enk}.  By combining a single mode cat state of the form
$|\sqrt{2} \alpha\rangle + |-\sqrt{2} \alpha \rangle$ with the vacuum
state on a 50/50 beam-splitter, the output state is of the form of a Bell
state in the $\{|\alpha\rangle,|-\alpha\rangle\}$ subspace:
\begin{eqnarray}\label{ent-cat-resource}
|\Psi\rangle&=& \frac{1}{\sqrt{\bar{\cal{N}}}} \left[|\alpha\rangle
|\alpha\rangle+ |-\alpha \rangle |-\alpha\rangle\right]
\end{eqnarray}
where $\bar{\cal{N}}$ is the normalization constant.
In this subspace we can also perform Bell-basis measurements by simply running the Bell
state creation in reverse: we interfere the two modes at a beam
splitter, then use photon counters to measure the number of photons in
each output mode \cite{kim01}.  We can then identify the four possible results:
(i) $(\mathrm{even},0)$, (ii) $(\mathrm{odd},0)$, (iii) $(0,\mathrm{even})$,
(iv) $(0,\mathrm{odd})$, where $(m,n)$ indicates counting $m$ and $n$
photons in the two modes respectively.  These results correspond to
each of the four Bell-cat states: (i) $\left(|-\alpha,-\alpha\rangle+
  |\alpha,\alpha\rangle\right)/\sqrt{2}$, (ii)
$\left(|-\alpha,-\alpha\rangle-|\alpha,\alpha\rangle\right)/\sqrt{2}$, (iii)
$\left(|-\alpha,\alpha\rangle+|\alpha,-\alpha\rangle\right)/\sqrt{2}$,
or (iv)
$\left(|-\alpha,\alpha\rangle-|\alpha,-\alpha\rangle\right)/\sqrt{2}$.
Note that there is also a fifth possibility of detecting zero photons
in both modes which indicates a failure of the
measurement. Fortunately, this occurs with probability of only $\sim
e^{-\alpha^2}$, and for $\alpha$ moderately large this is insignificant.

We can go further with entanglement and generate multi-mode entangled
states.  If a single mode cat state $|\alpha\rangle+|-\alpha\rangle$ is
input into one mode of an $N$ port symmetric beam-splitter with the
remaining input ports empty. The output state from this beam-splitter
is then the massively entangled GHZ-like state
\begin{equation}\label{nmodecat}
|\psi\rangle=\frac{1}{\sqrt{2}}\left[|\frac{\alpha}{\sqrt N},\frac{\alpha}{\sqrt N},\ldots, \frac{\alpha}{\sqrt N}\rangle+
|-\frac{\alpha}{\sqrt N},-\frac{\alpha}{\sqrt N},\ldots,-\frac{\alpha}{\sqrt N} \rangle\right].
\end{equation}

\section{Universal Quantum Logic Gates}
\label{sect:gates} 

The first application of this toolbox that we will review is a scheme
for quantum computation with coherent states \cite{ralph,ralph03}.
Consider an encoding of logical qubits in coherent states with the
logical 0 and 1 states being $|0\rangle_L = |-\alpha \rangle$ and
$|1\rangle_L = |\alpha\rangle$ respectively (An entirely equivalent
encoding would be $|0\rangle_L = |0 \rangle$ and $|1\rangle_L =
|2\alpha\rangle$ as discussed in \cite{ralph}, and these two encodings
are simply related by a displacement $D(-\alpha)$).  For convenience
and without loss of generality we will choose $\alpha$ to be real.
These qubits are not strictly orthogonal, but the approximation is
good for $\alpha$ even moderately large as $|\langle \alpha | -\alpha
\rangle|^{2} =e^{-4 \alpha^2}$.  For $\alpha \ge 2$ the overlap
between the zero and one logic qubit states is only $|\langle \alpha |
-\alpha \rangle|^{2} \le 10^{-6}$.

It is well known that one set of universal gates for qubits is comprised of
arbitrary single qubit rotations together with an entangling gate.
The single qubit rotations for our qubits can be built from four basic
single qubit gates. The first two gates are the bit and sign flip
operations and are given as follows:
\begin{itemize}
\item A bit-flip: The logical value of a qubit can be flipped by
  delaying it with respect to the local oscillator by half a cycle.
  Thus the ``bit-flip'' gate $X$ is given by $X=P(\pi)$. For example,
  $X(\mu|-\alpha \rangle + \nu |\alpha \rangle)=\mu|\alpha \rangle + \nu |-\alpha \rangle$.
 
\item A sign-flip: The sign flip gate $Z$ can be achieved using a
  teleportation protocol and the maximally entangled resource
  (\ref{ent-cat-resource}). Consider that we wish to sign flip the
  qubit $\mu|-\alpha \rangle + \nu |\alpha \rangle$. A Bell
  state measurement is performed between one half of the resource
  (\ref{ent-cat-resource}) and the qubit of interest.  Depending on
  which of the four possible outcomes are found the other half of the
  Bell state is projected into one of the following four states with
  equal probability: (i) $\mu|-\alpha \rangle + \nu |\alpha \rangle$,
  (ii) $\mu|-\alpha \rangle - \nu |\alpha \rangle$, (iii) 
  $\mu|\alpha \rangle + \nu |-\alpha \rangle$, and (iv) 
  $\mu|\alpha \rangle - \nu |-\alpha \rangle$.
  
  The bit flips in results three and four can be corrected using the
  $X$ gate above. After $X$ correction the gate has two possible outcomes:
  either the identity has been applied, in which case we repeat the
  process, or else the required transformation:
  \begin{equation}
  Z (\mu|-\alpha \rangle + \nu |\alpha \rangle) =
  \mu|-\alpha \rangle - \nu |\alpha \rangle.
  \end{equation}
\end{itemize}
The teleportation trick used in the $Z$ gate is incredibly useful and
can be used to `clean up' qubits that move slightly away from the 
$\{|-\alpha\rangle, |\alpha \rangle\}$ subspace \cite{ralph03}.
The remaining two operations are arbitrary rotations about the $Z$ and $X$
axis and like the sign flip operation $Z$ they also use a
teleportation protocol to achieve the gate.  These operations are
given by:
\begin{itemize}
\item An arbitrary rotation $\phi$ about the $Z$ axis, schematically
  depicted in figure~\ref{figsinglezcat} can be implemented by first
  displacing our arbitrary input qubit $\mu|-\alpha \rangle + \nu
  |\alpha \rangle$ by a small amount $\beta = \alpha \theta$ in the
  imaginary direction. This results in the state
  \begin{equation}
  \mu e^{-i \theta \alpha^{2}}|-\alpha(1-i \theta) \rangle + \nu e^{i \theta \alpha^{2}}|\alpha(1+i \theta) \rangle
  \end{equation}
  which is a small distance outside the computational space. The
  teleportation then projects us back into the qubit space resulting
  in the state
  \begin{equation}
  e^{-\theta^{2} \alpha^{2}/2}
  (e^{-i2 \theta \alpha^{2}}\mu|-\alpha \rangle + e^{i 2\theta \alpha^{2}}
  \nu |\alpha \rangle)
  \label{RZgate}
  \end{equation}
   \begin{figure}[!htb]
   \begin{center}
           \includegraphics[scale=0.45]{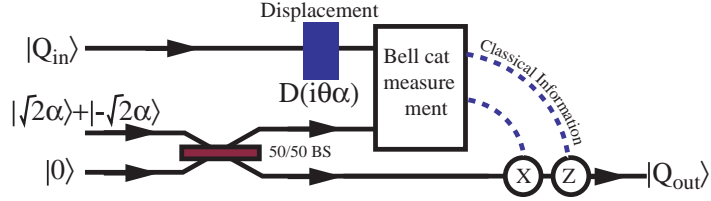}
   \end{center}
   \caption{Schematics for implementing a rotation around $Z$. We begin 
   by first shifting our qubit a small distance out of the computational basis 
   and then using teleportation to project back into the qubit space.}
   \label{figsinglezcat}
   \end{figure}
   This is a rotation around $Z$ by $4\theta\alpha^{2}$. This gate is
   near deterministic for a sufficiently small values of $\theta^{2}
   \alpha^{2}$. Repeated application of this gate can build up a
   finite rotation with high probability.
   
\item The fourth gate to consider is a rotation of $\pi/2$
   about the $X$ axis.  The gate is shown schematically in
   figure~\ref{figsinglexcat}.
   \begin{figure}
   \begin{center}
           \includegraphics[scale=0.45]{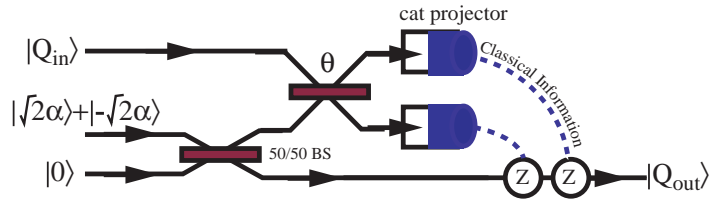}
   \end{center}
   \caption{Schematics for implementing a rotation of $\pi/2$ about the  $X$ axis. }
   \label{figsinglexcat}
   \end{figure}
   For an arbitrary input state $\mu |-\alpha \rangle + \nu |\alpha
   \rangle$, the  interaction $C_{a}C_{b}U_{BS}$ produces the output state
   (after correcting with $X$ and $Z$) 
   \begin{equation}
   e^{-\theta^2\alpha^2/4}\left[(e^{i \theta \alpha^{2}}\mu +
   e^{-i \theta \alpha^{2}}\nu) |-\alpha \rangle +
   (e^{-i \theta \alpha^{2}}\mu +
   e^{i \theta \alpha^{2}}\nu) |\alpha \rangle\right]
   \end{equation}
   where $C_{a}$ and $C_{b}$ represent cat state projections onto
   either the even or odd parity cat (i.e. photon counting and conditioning
   on even or odd number of photons). By choosing $2\theta\alpha^2=\pi/2$
   the gate will implement a $\pi/2$ rotattion around the $X$ axis.
\end{itemize}

By combining these gates it is possible to achieve an arbitrary single
qubit rotation.  If we can supplement these gates with a single two qubit
entangling operation between the qubits, then we have a universal set.
\begin{itemize}
\item We can implement an entangling gate in a similar way to the
  single qubit $Z$ rotation. A schematic circuit for the gate is depicted in
  figure~\ref{figtwozgate}.
\begin{figure}[!htb]
\begin{center}
        \includegraphics[scale=0.4]{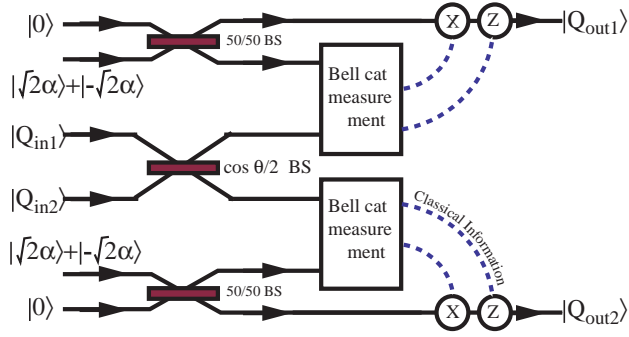}
\end{center}
\caption{Schematics of implementing an entangling gate. 
  For a sufficiently small value of $\theta^{2} \alpha^{2}$ this gate
  is near deterministic.  Repeated application of this gate can build
  up to a gate locally equivalent to a \textsc{cnot} gate, with high
  probability.}
\label{figtwozgate}
\end{figure}
If both our qubits are first mixed on a beam-splitter and are then
projected back into the qubit space of $\{|\pm \alpha\rangle\}$ using
teleportation, we find for an arbitrary input state $\nu|-\alpha
\rangle_{a} |-\alpha \rangle_{b}+ \mu|\alpha \rangle_{a} |-\alpha
\rangle_{b}+\tau|-\alpha \rangle_{a} |\alpha \rangle_{b}+\gamma|\alpha
\rangle_{a} |\alpha \rangle_{b}$ that the resultant state is
\begin{eqnarray}
 e^{i \theta \alpha^{2}}\nu|-\alpha \rangle_{a} |-\alpha
  \rangle_{b}&+&
e^{-i \theta \alpha^{2}}\mu|\alpha \rangle_{a} |-\alpha \rangle_{b} \nonumber \\
&+&
e^{-i \theta \alpha^{2}}\tau|-\alpha
\rangle_{a} |\alpha \rangle_{b}+ e^{i \theta \alpha^{2}}
\gamma|\alpha \rangle_{a} |\alpha \rangle_{b}\label{Ho1}
\end{eqnarray}
where, as before, we have assumed orthogonality of the qubit basis state and $\theta^{2}
\alpha^{2}\ll 1$. If we choose $2 \theta \alpha^{2}=\pi/2$ then
this gate will implement a \textsc{cnot} up to single-qubit rotations \cite{ralph03}.
\end{itemize}
This then completes a universal set of gates. In reference
\cite{ralph03} details are given on using further nested teleportation
to make these gates deterministic without requiring
$\theta^{2}\alpha^{2}<<1$. These gates can be used for both quantum
computation and communication.

\section{Quantum Metrology}
\label{sect:metrology}

In this section we illustrate the utility of the Schr\"odinger cat
states for two metrology applications --- the detection of weak
forces, and high precision phase measurements.

\subsection{The detection of weak forces}

Before we begin our discussion of the application of Schr\"odinger cats
states to weak force detection \cite{munro}, it is essential to establish the best
classical limit. It is well known that when a classical force given by
$F(t)$ acts for a fixed time on a simple harmonic oscillator, it
displaces the complex amplitude of this oscillator in phase space. The
resulting amplitude and phase of the displacement are determined by
the time dependence of the force \cite{braginsky92}.  If the
oscillator begins in a coherent state $|\alpha_0\rangle$ (with
$\alpha_0$ real) then a displacement $D(i\epsilon)$ (assumed for
simplicity to be orthogonal to the coherent amplitude of the initial
state) causes the coherent state to evolve to $e^{i \epsilon
  \alpha_0}|\alpha_0+i\epsilon\rangle$.  The maximum signal to noise
ratio is then $SNR=S/\sqrt{V}=2\epsilon$. This must be greater than
unity for the displacement to be resolved and hence this establishes the
standard quantum limit (SQL) \cite{simple} of $\epsilon_{SQL}\geq
1/2$.

\begin{figure}[!htb]
\begin{center}
\includegraphics[scale=0.4]{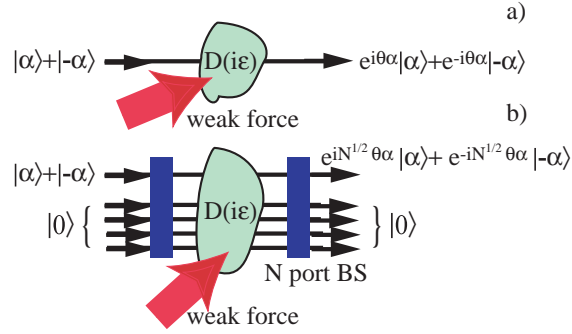}
\end{center}
\caption{Schematic diagram of the action of a weak force causing a displacement 
$D(i \epsilon)$ on a Schr\"odinger cats state  $|\alpha\rangle+|-\alpha\rangle$. In a) 
a single mode case is illustrated while in b) an N mode situation is considered.}
\label{figweakforce}
\end{figure}

It is also well known that this limit may be overcome if the
oscillator is prepared in a non-classical state. However, what is the
sensitivity achieved by (\ref{eqn:cat}), and does this reach the
ultimate (Heisenberg) limit? When a weak classical force acts on the
even photon number cat state $|\alpha\rangle+|-\alpha\rangle$ with
$\alpha$ real (see figure~\ref{figweakforce}a) it is displaced to
\begin{eqnarray}\label{displacedcat}
|\phi\rangle_{out}&\approx& \frac{1}{\sqrt{2}}\left(e^{i \epsilon \alpha}|\alpha\rangle+e^{-i
\epsilon \alpha}|-\alpha\rangle\right ).
\end{eqnarray}
Our problem is thus reduced to finding the optimal readout to be able
to distinguish (\ref{displacedcat}) from
$|\alpha\rangle+|-\alpha\rangle$.  The theory of optimal parameter
estimation \cite{bcm} indicates that the limit on the precision with
which the parameter $\epsilon \alpha$ can be determined is
$(\delta\theta)^2\geq 1/{{\rm Var}(\hat{\sigma}_x)_{in}}$ where
${\rm Var}(\hat{\sigma}_x)_{in}$ is the variance in the generator of the
rotation in the input state $|\alpha\rangle+|-\alpha\rangle$. In this
case the variance is simply unity. It thus follows that the minimum
detectable force is
%\begin{equation}
$\epsilon \geq {1}/{2\sqrt{\bar n}}$
%\end{equation}
where $\bar n$ is the mean photon number given by $\bar n=|\alpha|^2$.
It is straight forward to show this `measurement' is the Heisenberg
limit for a displacement measurement. An interesting question is what
type of measurement is required to achieve this limit. In effect we
need to be able to distinguish the even parity cat state from the odd
parity cat state. Currently this is experimentally challenging.
However by performing a Hadamard operation (one of the single qubit
gates discussed previously), the even and odd Schr\"odinger cats are
transformed to the coherent states $|\alpha\rangle$ from
$|-\alpha\rangle$ which can be easily distinguished via a standard
homodyne measurement.

If the weak force acts over a reasonable spatial range it would be
possible to have a number of spatial modes of light being affected.
Could this help us exceed the limit above, even if we constrainted the
total mean photon number of the entire multimode system? We depict in
figure~\ref{figweakforce}b a schematic for the setup of a proposed
experiment. Using a single mode cat state and an $N$ port symmetric beam-splitter
we can generate the state~(\ref{nmodecat}), 
which has a total mean photon number of $n_{tot}=|\alpha|^2$. We now
assume that the weak force acts simultaneously on all modes of this
$N$ party entangled state, displacing them each by an amount $D(i
\epsilon)$ (for $\epsilon \ll 1$). The resulting state after the
action of the force is
\begin{equation}\label{fulldisplaced}
|\psi (\theta)\rangle=\frac{1}{\sqrt{2}} \left[ e^{i \sqrt{N} \epsilon \alpha} 
|\frac{\alpha}{\sqrt N},\ldots, \frac{\alpha}{\sqrt N}\rangle  +
e^{-i \sqrt{N} \epsilon \alpha}|-\frac{\alpha}{\sqrt N},\ldots,-\frac{\alpha}{\sqrt N} \rangle 
\right],
\end{equation}
where we have neglected the small displacement that occurs to the
coherent state.  The theory of optimal parameter estimation indicates
that the limit on the precision with which the displacement parameter
$\epsilon$ be estimated is bounded by
\begin{equation}
\epsilon_{min}=\frac{1}{\sqrt{N \left[1+4 n_{tot}\right]}}\sim \frac{1}{2\sqrt{N n_{tot}}} \nonumber
\end{equation}
for $n_{tot}\gg 1$. If however we had used $N$ independent cat states
each with a photon number $n_{tot}/N$ then $\epsilon_{min}$ for
the entire system would have scaled as $\epsilon_{min}\sim 1/\sqrt{4
  n_{tot}}$ which is the same result we obtained for the single mode
case. For large $n_{tot}$, the preferred regime to work in, we find
that the $N$ mode entangled situation gives an extra $\sqrt{N}$
improvement over the single mode cat situation for the same total mean
photon number. Now how do we interpret such results?  The effect that
we are seeing is due to the weak force acting equally on all $N$ modes
and the state between the $N$ port beam-splitters being highly
entangled. Does this result in a violation of the Heisenberg limit of
$1/ \sqrt{n_{tot}}$ which we previously mentioned? The answer is no. A
careful analysis using parameter estimation of this multimode
situation indicates that our result is at the Heisenberg limit. For
displacement measurements the Heisenberg limit does depend on the
number of modes.

These results indicate that subject to the spatial bandwidth of the weak
classical force it seems optimal for a cat state with fixed mean
photon $n_{tot}$ to be split and entangled over as many modes as
feasible. This in the absence of loss gives the best sensitivity. Such
techniques are likely to work for other non classical continuous
variable states.

\subsection{High precision phase measurements}
The second metrological example we are going to investigate is the estimation of phase.
The classic situation to consider is Ramsey fringe interferometry which was 
first introduced by Bollinger et al. \cite{bollinger} in the mid nineties. 
In Ramsey fringe interferometry the objective is to
detect the relative phase difference between two superposed qubit basis states 
$|0\rangle$ and $|1\rangle$. This phase difference problem reduces to a 
quantum parameter estimation situation in which a  
unitary transformation $U(\theta)=\exp[i\theta \hat{Z}]$
(with $\hat{Z}=|1\rangle\langle 1|-|0\rangle\langle 0|$) induces a relative phase in the
specified basis. For example, an initial state of the form 
$c_0 |0\rangle+c_1 |1\rangle$ evolves to $c_0 e^{-i\theta} |0\rangle+c_1 e^{i\theta} |1\rangle$ 
under the above unitary operation. When can we distinguish these two states? Is there an 
optimal choice of initial state?
The theory of quantum parameter estimation \cite{bcm} indicates for this situation 
that we should choose the initial state as
$|\psi\rangle_i=(|0\rangle+|1\rangle)/\sqrt{2}$ and that the optimal
measurement is a projective measurement in the basis
$|\pm\rangle=|0\rangle\pm|1\rangle$. The probability of obtaining the result
$+$ is  $P(+|\theta)=\cos^2\theta$. For $N$ repetitions of this measurement the
uncertainty in the inferred parameter $\theta$ is $\delta\theta=1/\sqrt{N}$. This is known as 
the standard quantum limit. It was noted by Bollinger et al. \cite{bollinger}  that
a more effective way to use the $N$ two level systems is to
first prepare them in the maximally entangled state,
\begin{equation}
|\psi\rangle=\frac{1}{\sqrt{2}}(|0\rangle_1|0\rangle_2\ldots|0\rangle_N+|1\rangle
_1|1\rangle_2\ldots|1\rangle_N)
\label{entangle}
\end{equation}
and then subject the entire state to the unitary transformation 
$U(\theta)=\prod_{i=1}^N\exp(-i\theta\hat{Z}_i)$. After the unitary transformation the state (\ref{entangle}) 
evolves to 
\begin{equation}
|\psi\rangle=\frac{1}{\sqrt{2}}(\exp(-i N \theta)|0\rangle_1|0\rangle_2\ldots|0\rangle_N
+\exp(i N \theta)|1\rangle_1|1\rangle_2\ldots|1\rangle_N)
\end{equation}
The uncertainty in the estimation of the parameter $\theta$ then achieves
the Heisenberg lower limit of $\delta\theta=1/N$. This would seem to
indicate, as in the weak force case, that entanglement is a critical
requirement to achieve the improved sensitivity. Let us examine this
point a little further for the phase estimation situation. The Hilbert
space of $N$ two level systems is a tensor product space of dimension
$2^N$.  The entangled state given in equation~(\ref{entangle}) however
resides in a much smaller $N+1$ dimensional irreducible subspace of
permutation symmetric states \cite{dicke}. We may use an SU(2)
representation to write the entangled state
$0\rangle_1|0\rangle_2\ldots|0\rangle_N+|1\rangle
_1|1\rangle_2\ldots|1\rangle_N$ in the form
\begin{equation}
|\psi\rangle=\frac{1}{\sqrt{2}}(|-N/2\rangle_{N/2}+|N/2\rangle_{N/2}).
\end{equation}
This is just an SU(2) `cat state' for $N$ two-level atoms. Hence a single
$N$ level atom can achieve the same phase sensitivity as a maximally 
entangled GHZ state since it can be written in 
the form $|-N/2\rangle_{N/2}+|N/2\rangle_{N/2}$.  This would also seem to 
indicate that a superposition of coherent states (a cat state) can provide the same 
phase resolution. 
\begin{figure}[!htb]
\begin{center}
\includegraphics[scale=0.5]{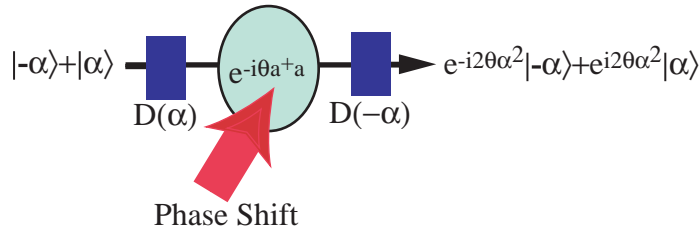}
\end{center}
\caption{Schematics of quantum circuit illustrating how a phase shift can be seen 
on an input state of the form}
\label{figphase}
\end{figure}
In figure~\ref{figphase} a schematic diagram is shown for the cat state yielding the
Heisenberg limited phase resolution. Such phase shifts could be used to
resolve precisely very small length intervals, a {\it quantum ruler} \cite{ralphcat} in effect. 
As $\alpha$ increases, a number of high visibility, narrowly spaced fringes emerge, which could
enable very short length intervals to be accurately measured. As an
example suppose our laser wavelength is 10$\mu m$. In a standard
interferometer this would enable length intervals of $5 \mu m$ to
be stepped off. However using the cat-state technique with an $\alpha$ of 10 
leads to the fringe separation being reduced to $1 \mu m$.

The preceding disscussion shows that entanglement is not necessary to
achieve a Heisenberg limited phase measurement. What entanglement
allows however, is to create an effective cat state without the need
to resort to creating a superposition between the ground state and a
highly excited one.

\section{Concluding Remarks}

In the paper we have presented a toolbox of techniques that make use
of superpositions of coherent states. Using this toolbox we have
presented a quantum computation scheme based on encoding qubits as
coherent states, and their superposition. The optical networks
required are conceptually simple and require only linear interactions,
homodyne measurements and photon counting.  We have concentrated on
the simplest implementation which unfortunately requires large
$\alpha$. However with a modest increase in complexity the
non-deterministic operation of the gates at low $\alpha$ can form the
basis of a scalable system \cite{ralph03}.  We have also shown how the same
toolbox can be used to achieve extremely sensitive force detection and
phase measurements.  

An open and very interesting question is whether the toolbox of
techniques and states we have described can be transfered to other
systems where we can formulate coherent states, for example SQUIDs. In
those systems it may be possible to augment or simplify the toolbox
making the quantum information applications more accessible.

%%%%%%%%%%%%%%%%%%%%%%%%%%%%%%%%%%%%%%%%%%%%%%%%%%%%%%%%%%%%%
\ack          

This work was supported in part by the European project RAMBOQ, 
the SUMITOMO foundation, EPSRC and the Australian Research Council. 
AG acknowledges support form the New Zealand Foundation for Research, Science 
and Technology under grant UQSL0001. SLB currently holds a Royal Society 
Wolfson Research Merit Award. We thank Tim Spiller and Ray Beausoleil for valuable
discussions. \\

%%%%%%%%%%%%%%%%%%%%%%%%%%%%%%%%%%%%%%%%%%%%%%%%%%%%%%%
%%%%% References %%%%%


\begin{thebibliography}{99}

\bibitem{simple} See for instance: D.F.~Walls and G.J.~Milburn, {\it Quantum Optics}
(Springer-Verlag, Berlin, 1994).  
%
\bibitem{nielsen} M.~Nielsen and I.~Chuang,
{\it Quantum computation and quantum information}
(Cambridge University Press, Cambridge, UK 2000).
%
\bibitem{mil88} G.J.~Milburn, Phys. Rev. Lett. {\bf 62}, 2124 (1989).
%
\bibitem{KLM} E.~Knill and L.~Laflamme and G.J.~Milburn,
Nature {\bf 409}, 46 (2001).
%
\bibitem{pre} D.~Gottesman, A.~Kitaev, J.~Preskill, Phys Rev A {\bf 64},
012310 (2001).
%
\bibitem{teleport} A.~Furasawa, J.L.~S{\o}rensen, S.L.~Braunstein, C.A.~Fuchs, 
H.J.~Kimble and E.S.~Polzik, Science {\bf 282}, 706 (1998).
%
\bibitem{lloyd} S.~Lloyd, S.L.~Braunstein, Phys Rev Lett {\bf 82}, 1784 (1999).
%
\bibitem{sanders} S.D.~Bartlett, Hubert de Guise, B.C.~Sanders, Phys. Rev. A {\bf 65}, 052316 (2002).
%
\bibitem{kim01} H.~Jeong and M.S.~Kim, Phys. Rev. A {\bf 65}, 042305 (2002).
%
\bibitem{ralph} T.C.~Ralph, W.J.~Munro and G.J.~Milburn, Quantum Computation with Coherent States, 
Linear Interactions and Superposed Resources, Hewlett Packard Labs Tech Report 2001-266.
%
\bibitem{ralph03} T.C.~Ralph, A.~Gilchrist, G.J.~Milburn, W.J.~Munro and S.~Glancy, Phys. Rev. A,
  \textbf{68}, 042319 (2003).
%
\bibitem{Polzik Carry and Kimble 92} E.S.~Polzik, J.~Carry, and H.J.~Kimble, Phys. Rev. Lett. 
{\bf 68}, 3020 (1992).
%
\bibitem{99kim902}
J.~Kim, S.~Takeuchi, Y.~Yamamoto, and H.H.~Hogue, Appl. Phys. Lett. {\bf 74},
  902  (1999).
%
\bibitem{99takeuchi1063}
S.~Takeuchi, J.~Kim, Y.~Yamamoto, and H.H. Hogue, Appl. Phys. Lett. {\bf 74},
  1063  (1999).
%
\bibitem{van enk} S.J.~van Enk and O.~Hirota, Phys. Rev. A {\bf 64}, 022313, (2001).
%
\bibitem{caves} C.M.~Caves, K.S.~Thorne, R.W.P.~Drever, V.D.~Sandberg and M.~Zimmermann, Rev. Mod. Phys. {\bf 52}, 341, (1980).
%
\bibitem{munro} W.J.~Munro, K.~Nemoto, G.J.~Milburn and S.L.~Braunstein, Phys. Rev. A {\bf 66}, 023819 (2002).
%
\bibitem{bollinger} J.J.~Bollinger, Wayne~M.~Itano, D.J.~Wineland and D.J.~Heinzen, Phys. Rev. A {\bf 54}, R4649, (1996).
%
\bibitem{huelga}  S.F.~Huelga, C.~Macchiavello, T.~Pellizzari, A.K.~Ekert,
M.B.~Plenio and J.I.~Cirac, Phys. Rev. Lett. {\bf 79}, 3865, (1997).
%
\bibitem{dakna97} M.~Dakna, T.~Anhut, T.~Opatrny, L.~Kn\"oll and
D.-G.~Welsch, Phys. Rev. A {\bf 55}, 3184 (1997).
%
\bibitem{song} S.~Song, C.~M.~Caves and B.~Yurke,
Phys. Rev. A {\bf 41}, 5261 (1990).
%
\bibitem{braginsky92} V.~Braginsky and F.~Ya.~Khalili, {\em Quantum
measurement}, (Cambridge University Press, Cambridge, 1992).
%
\bibitem{bcm}S.L.~Braunstein , C.M.~Caves and G.J.~Milburn, Annals of
Physics {\bf 247}, 135, (1996).
%
\bibitem{dicke} R.H.~Dicke, Phys. Rev. {\bf 93}, 99 (1954).
%
\bibitem{ralphcat}  T.C.~Ralph, Phys. Rev. A {\bf 65}, 042313 (2002).


\end{thebibliography}
\end{document}